\documentclass[aps,prl,twocolumn,superscriptaddress]{revtex4-1}
\usepackage[dvips]{epsfig}

\usepackage[sort&compress]{natbib}

\begin{document}

%\preprint{}

\title{Telecommunications-band heralded single photons from a silicon nanophotonic chip}



\author{Marcelo Davan\c{c}o}
\affiliation{Center for Nanoscale Science and Technology, National
Institute of Standards and Technology, Gaithersburg, MD 20899, USA}
\affiliation{Maryland NanoCenter, University of Maryland, College
Park, MD 20742, USA}
\author{Jun Rong Ong}
\affiliation{University of California, San Diego, Mail Code 0407, La
Jolla, California 92093}
\author{Andrea Bahgat Shehata}
\affiliation{Politecnico di Milano, Dipartimento di Elettronica e
Informazione, Piazza Leonardo da Vinci 32, 20133 Milano, Italy}
\author{Alberto Tosi}
\affiliation{Politecnico di Milano, Dipartimento di Elettronica e
Informazione, Piazza Leonardo da Vinci 32, 20133 Milano, Italy}
\author{Imad Agha}
\affiliation{Center for Nanoscale Science and Technology, National
Institute of Standards and Technology, Gaithersburg, MD 20899, USA}
\affiliation{Maryland NanoCenter, University of Maryland, College
Park, MD 20742, USA}
\author{Solomon Assefa}
\affiliation{IBM Thomas J. Watson Research Center, Yorktown Heights,
New York 10598}
\author{Fengnian Xia}
\affiliation{IBM Thomas J. Watson Research Center, Yorktown Heights,
New York 10598}
\author{William M.J. Green}
\affiliation{IBM Thomas J. Watson Research Center, Yorktown Heights,
New York 10598}
\author{Shayan Mookherjea}
\email{smookherjea@ucsd.edu} \affiliation{University of California,
San Diego, Mail Code 0407, La Jolla, California 92093}
\author{Kartik Srinivasan}
\email{kartik.srinivasan@nist.gov} \affiliation{Center for Nanoscale
Science and Technology, National Institute of Standards and
Technology, Gaithersburg, MD 20899, USA}


\date{\today}

\begin{abstract}
We demonstrate heralded single photon generation in a
CMOS-compatible silicon nanophotonic device. The strong modal
confinement and slow group velocity provided by a coupled resonator
optical waveguide (CROW) produced a large four-wave-mixing
nonlinearity coefficient $\gamma_{\text{eff}}\approx4100$
W$^{-1}$m$^{-1}$ at telecommunications wavelengths. Spontaneous
four-wave-mixing using a degenerate pump beam at 1549.6 nm created
photon pairs at 1529.5 nm and 1570.5 nm with a
coincidence-to-accidental ratio exceeding 20. A photon correlation
measurement of the signal (1529.5 nm) photons heralded by the
detection of the idler (1570.5 nm) photons showed antibunching with
$g^{(2)}(0)=0.19\pm0.03$. The demonstration of a single photon
source within a silicon platform holds promise for future integrated
quantum photonic circuits.

\end{abstract}

\pacs{}
%\keywords{}

\maketitle

Chip-based single photon sources leverage the scalability and device
integration afforded by modern semiconductor fabrication technology
for quantum information processing
applications~\cite{ref:OBrien_Furusawa_Vuckovic,ref:Shields_NPhot}.
There are two dominant approaches to single photon generation at
optical wavelengths.  The first is through radiative decay of a
single quantum emitter such as an atom or quantum dot that is
"triggered" by excitation pulses.  The second is through spontaneous
photon pair production, in which the detection of one photon of the
pair provides the time stamp by which the remaining ("heralded")
single photon is identified.  Both approaches for single photon
generation were first demonstrated in bulk optical systems decades
ago~\cite{ref:Kimble_Mandel_PRL,ref:Grangier_single_photon,ref:Hong_Mandel_PRL}.
Since then, chip-based triggered single photon sources based on
systems such as a single quantum dot in a
nanocavity~\cite{ref:Michler,ref:Santori,ref:Shields_NPhot} have
been widely studied, but this work is generally in
cryogenically-cooled III-V semiconductor
systems~\cite{ref:Michler_book}. In contrast, photon pair production
and subsequent heralded single photon generation, which are usually
based on second- and third-order nonlinear processes that are
achievable in a broader class of materials and at room temperature,
have primarily been studied in larger systems such as bulk
crystals~\cite{ref:Fasel_Zbinden_NJP,ref:Mosely_Walmsley_PRL},
periodically-poled
waveguides~\cite{ref:Uren_Walmsley_PRL,ref:Alibart_Tanzilli_OL}, and
optical
fibers~\cite{ref:Voss_Kumar_PTL,ref:Rarity_OE,ref:Goldschmidt_Migdall_PRA,ref:Cohen_Walmsley_PRL}.
Recently, however, researchers have begun exploring four-wave-mixing
(FWM) and photon pair production in CMOS-compatible silicon
nanophotonic
devices~\cite{ref:Sharping_Kumar_OE,ref:Harada_Itabashi_OE,ref:Clemmen_Massar_OE,ref:Xiong_PCWG_CAR},
which support a strong third-order optical nonlinearity and have the
potential for significant levels of integration with other quantum
optical components. Here, we build upon this work by demonstrating
not only photon pair production, but also heralded single photon
generation in a chip-based, silicon nanophotonic device operating in
the telecommunications band and at room temperature.

\begin{figure*}[t]
\centering
\includegraphics[width=\linewidth, clip=true]{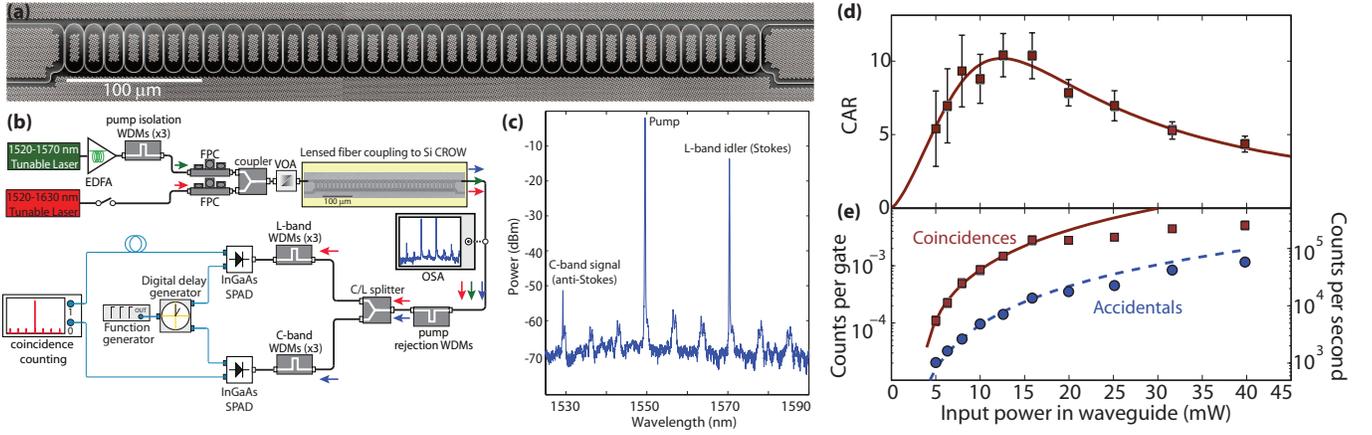}
\caption{Photon pair production in a silicon CROW. (a) Scanning
electron microscope image of the 35-ring CROW used in this work. (b)
Schematic of the experimental setup used to measure correlated
photon pairs generated by the CROW.  The 1520 nm to 1630 nm tunable
laser is used for stimulated FWM experiments to identify the signal
and idler wavelengths, but is disconnected during SFWM/photon pair
generation measurements. EDFA~=~erbium-doped fiber amplifier,
WDM~=~wavelength division multiplexer, FPC~=~fiber polarization
controller, VOA~=~variable optical attenuator, OSA~=~optical
spectrum analyzer, SPAD~=~single photon avalanche diode. (c) FWM
spectrum in which a 1549.6 nm pump amplifies a 1570.5 nm probe and
generates a new field at 1529.5 nm.  The spectral peaks in-between
the pump and signal/idler fields are due to transmission of
(unfiltered) EDFA spontaneous emission (ASE) through the CROW
passbands. In photon pair measurements, this ASE is suppressed by
$>150$ dB by the pump isolation WDMs. (d) Coincidence-to-accidental
ratio (CAR) as a function of power at the CROW input, for continuous
wave pumping~[26]. (e) Number of coincidences (red) and accidentals
(blue) at the CROW output as a function of power at the CROW input.
Results are plotted in units of (left y-axis) counts per gate and
(right y-axis) counts per second~[27].} \label{fig:fig1}
\end{figure*}

Since silicon lacks a second-order optical nonlinearity, photon pair
production uses the third-order (ultrafast Kerr) nonlinearity,
typically in the degenerate four-wave-mixing configuration where a
single pump beam at frequency $\omega_{p}$ generates photons at
signal ($\omega_{s}$) and idler ($\omega_{i}$) frequencies, with
energy conservation requiring
2$\omega_{p}$=$\omega_{s}$+$\omega_{i}$ and momentum conservation
(phase-matching) being a requirement for appreciable pair
production~\cite{ref:Agrawal_NFO}.  Silicon nanophotonic waveguides
have an effective nonlinearity coefficient
$\gamma_{\text{eff}}\approx200$ W$^{-1}$m$^{-1}$ that is four orders
of magnitude larger than that of highly nonlinear optical
fiber~\cite{ref:Sharping_Kumar_OE,ref:Harada_Itabashi_OE}. Also,
spontaneous Raman scattering, a broadband noise source in optical
fibers that can require them to be
cryogenically-cooled~\cite{ref:Lee_Kumar_OL}, is generally less
important in silicon, where it is narrowband and can thus be more
easily avoided.  On the other hand, in comparison to silica fibers,
silicon devices exhibit two-photon absorption (TPA) and free-carrier
absorption (FCA) at higher pump powers, incur coupling losses if
photons have to be coupled to input/output optical fibers, and are
generally limited to a few centimeters of waveguide length on a
chip.

Our device geometry is a silicon coupled-resonator optical-waveguide
(CROW) as shown in Fig.~\ref{fig:fig1}(a). The CROW consists of
$N=35$ directly-coupled microring resonators (loss~=~0.21 dB/ring),
such that each eigenmode is a collective resonance of all $N$
resonators.  Light is transmitted through the CROW in a
disorder-tolerant slow light regime, with slowing factor $S =
c/v_{g}$ between 5 and 12, depending on the wavelength ($c$ is the
speed of light and $v_g$ is the group velocity).  As
$\gamma_{\text{eff}}$ is enhanced by a factor $S^2$, the CROW
achieves higher levels of conversion within the limited footprint
available on a chip. Indeed, in ref.~\cite{ref:Ong_Mookherjea_OL},
we have shown classical FWM with $\gamma_{\text{eff}}\approx
4100$~W$^{-1}$m$^{-1}$, representing +16~dB enhanced conversion
compared to a conventional nanophotonic waveguide, for over $>10$
THz (80 nm) separation between signal and idler. This and other
reports of FWM in CROWs~\cite{ref:Morichetti_Melloni_NComm} have
shown similar conversion efficiencies to the best photonic crystal
waveguides (PCWGs).  However, such widely-separated wavelengths,
which span a significant fraction of the fiber-optic
telecommunications window, are difficult to achieve in PCWGs because
of the limited bandwidth of their slow-light regime compared to
CROWs; $\approx$1.25 THz (10 nm) signal-idler separation was
reported in ref.~\cite{ref:Xiong_PCWG_CAR}.

%CROWs also compete favorably with single
%resonators~\cite{ref:Clemmen_Massar_OE}: whereas a single high-$Q$
%resonator is difficult to fabricate so precisely as to display
%matched resonances to desired pump, signal, and idler wavelengths,
%CROWs demonstrate flat-top passbands with zero dispersion near the
%center of each band. CROWs are simultaneously nonlinear mixers and
%optical filters, and also show greater out-of-band filtering with
%steeper filter sidewalls than possible with a single microring.
%Finally, since FWM conversion efficiency scales quadratically with
%the number of resonators, significant benefits can be achieved when
%fabrication disorder can be overcome to the extent that dozens of
%microrings can be coupled, as we have shown
%recently~\cite{ref:Cooper_Mookherjea_OE}.

We first show photon pair production from the Si CROW device, using
the experimental setup depicted in Fig.~\ref{fig:fig1}(b).
Time-correlated signal and idler photons are expected to be
generated in multiple pairs of CROW transmission bands that are
approximately equally red- and blue-detuned from our amplified pump
beam at 1549.6 nm, as demonstrated in previous classical FWM mixing
experiments~\cite{ref:Ong_Mookherjea_OL}.  We choose a signal-idler
pair at 1529.5 nm and 1570.5 nm, as shown in Fig.~\ref{fig:fig1}(c).
Here, to show the classical FWM process, a strong pump at 1549.6 nm
was combined with a probe field at 1570.5 nm, resulting in the
addition of stimulated photons into the 1570.5 nm field and
generation of a new field at 1529.5 nm.  For spontaneous FWM (SFWM)
experiments, the 1570.5 nm probe field was disconnected so that
spontaneous photons are generated in the signal and idler bands. The
1549.6 nm pump was filtered to a 1.0 nm bandwidth through cascaded
WDM and tunable filters, and light was coupled to and from the chip
(loss~=~5 dB per coupler) using tapered lensed fibers and polymeric
overlaid waveguide couplers. Output light from the chip was filtered
by a set of WDM pump-rejection filters (120 dB estimated pump
rejection at 1550 nm $\pm$ 3 nm) and then routed through cascaded C-
and L- band WDM filters (estimated 150 dB pump isolation; 0.5 nm
bandwidth) to spectrally separate and isolate the signal and idler
photons, respectively. The signal (C-band) and idler (L-band)
photons were detected by InGaAs/InP Single-Photon Avalanche Diodes
(SPADs)~\cite{ref:Ribordy} gated electronically at 1 MHz (10~$\%$
detection efficiency, 20 ns gate width, and 10 $\mu$s dead-time),
and raw coincidences ($C_{\text{raw}}$) and accidentals
($A_{\text{raw}}$) were measured by a time-correlated single photon
counting (TCSPC) system operating with 512 ps timing resolution,
with typical measurement integration times between 1800 s and 5400
s. Coincidences due to dark counts ($D$) were measured separately
for both integration times at each detector, and subtracted to yield
$C=C_{\text{raw}}-A_{\text{raw}}$ and $A=A_{\text{raw}}-D$, with the
coincidence-to-accidental ratio given as
$\text{CAR}=C/A$~\footnote{Raw coincidences $C_{\text{raw}}$ are
counted over a 512 ps bin at zero time delay between the C$-$band
and L$-$band paths. Raw accidentals ($A_{\text{raw}}$) are taken as
the average over thirty separate 512 ps bins at time delays of $\{1
{\mu}s, 2 {\mu}s,...,30 {\mu}s \}$, corresponding to the 1 MHz
trigger rate, with coincidences due to dark counts $D$ determined in
the same way. The uncertainties in $A_{\text{raw}}$ and $D$ are one
standard deviation values and are propagated to generate the error
bars in the CAR plot.}.

\begin{figure*}
\includegraphics[width=\linewidth, clip=true]{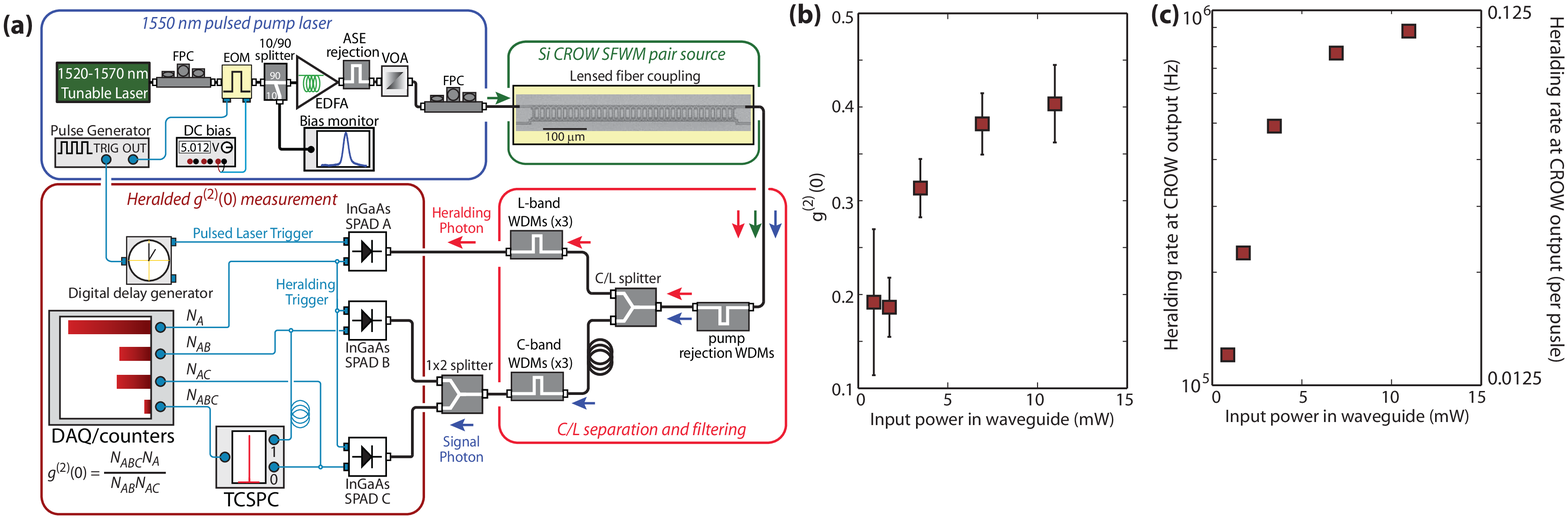}
\caption{Heralded single photon measurement. (a) Schematic of the
experimental setup used to perform heralded single photon
measurements.  The Si CROW waveguide is pumped by a pulsed 1549.6 nm
laser (2.5 ns pulses, 8 MHz repetition rate) generated by a
modulated and amplified diode laser.  Generated photon pairs are
spectrally isolated and separated into the C-band (1529.5 nm) and
L-band (1570.5 nm).  Detection of an L-band photon by an InGaAs/InP
SPAD is used to trigger a Hanbury-Brown and Twiss photon correlation
measurement on the C-band photon. (b) Heralded $g^{(2)}(0)$ as a
function of average power at the CROW input~[30]. (c) Heralding rate
at the CROW output as a function of average power at the CROW input.
Results are plotted in units of (left y-axis) heralding photons per
second and (right y-axis) heralding photons per pulse.}
\label{fig:fig3}
\end{figure*}

CAR under continuous wave (cw) excitation is shown in
Fig.~\ref{fig:fig1}(d) as a function of the input power into the
CROW.  CAR initially increased and then rolled off at higher
intensities, which is the anticipated behavior based on other
studies~\cite{ref:Sharping_Kumar_OE,ref:Harada_Itabashi_OE,ref:Xiong_PCWG_CAR,ref:Clemmen_Massar_OE},
where at low powers CAR is thought to be limited by detector noise,
while at higher powers, nonlinear loss and multiple pair generation
are the limiting factors.  Peak CAR was 10.4 $\pm$ 1.4 at an input
power of 12 dBm, which was below the level for 1 dB excess nonlinear
absorption in these CROWs~\cite{ref:Ong_Mookherjea_OL}. In
Fig.~\ref{fig:fig1}(e), we plot the coincidence and accidental rates
at the output of the CROW~\footnote{Coincidence and accidental rates
at the CROW output are determined by taking the measured values and
accounting for detector efficiency, filter losses, and output
coupling loss from the CROW into the lensed fiber.}. At peak CAR,
the coincidence rate is $\approx1.5{\times}10^{-3}$ per detector
gate; considering the cw pumping and the 1 MHz detector trigger rate
and 20 ns gate width, this corresponds to a pair coincidence rate of
$\approx$73 kHz. Figure~\ref{fig:fig1}(e) also shows quadratic fits
(solid lines) to the six lowest power data points; the sub-quadratic
dependence of $C$ and $A$ at higher pump powers was most likely
related to TPA/FCA effects.  We compared the pair production
performance of our CROW with a conventional single mode nanophotonic
silicon wire waveguide (length of 2.63 cm, loss = 2.6 dB/cm,
coupling loss = 5 dB per coupler) on the same chip. A peak CAR of
8.5 $\pm$ 1.0 was measured for this device, with a pair coincidence
rate of 95 kHz. Thus, the CROW photon pair source moderately
outperformed a conventional silicon waveguide whose physical
footprint was 54 times longer.

We next consider heralded single photon generation from this device
(Fig.~\ref{fig:fig3}(a)). Here, the detection of an L-band idler
photon indicates (heralds) the presence of its twin, and a photon
correlation measurement on these heralded photons confirms their
single photon
character~\cite{ref:Grangier_single_photon,ref:Hong_Mandel_PRL}. We
pumped the CROW using a pulsed source, which was created by
modulating and amplifying a tunable diode laser at 1549.6 nm to
create 2.5 ns wide, 8 MHz repetition rate pulses. C-band signal and
L-band idler photons were spectrally separated and isolated in the
same way as above, but now the C-band signal photons were split by a
50/50 coupler, with each C-band path detected by an InGaAs/InP SPAD
(20~$\%$ detection efficiency, 20 ns gate width, and no deadtime).
The detectors in this Hanbury-Brown and Twiss (HBT) photon
correlation measurement setup (labeled SPAD B and SPAD C in
Fig.~\ref{fig:fig3}(a)) were triggered by the detection of an L-band
idler photon (the herald). The L-band photons were detected by a
high-performance InGaAs/InP SPAD~\cite{ref:Tosi}, labeled SPAD A in
Fig.~\ref{fig:fig3}(a), which operates at 30~$\%$ detection
efficiency, 10 ns gate width, and 10 $\mu$s dead time and is
triggered at 8 MHz by the electro-optic modulator driver.  The
normalized value of the photon correlation measurement on the C-band
signal photons at zero time delay, $g^{(2)}(0)$, is given by
$g^{(2)}(0)=\frac{N_{ABC}N_{A}}{N_{AB}N_{AC}}$~\cite{ref:Beck_heralding}.
Triple coincidences $N_{ABC}$, corresponding to simultaneous events
on all three detectors, were recorded over a 2.5 ns bin using the
TCSPC. Double coincidences $N_{AB}$ and $N_{AC}$, corresponding to
simultaneous events on SPADs A and B or SPADs A and C, were given by
the photon detection rates on SPAD B and SPAD C.  The number of
heralding photons $N_{A}$ is determined by the detection rate on
SPAD A, and a typical integration time of 1500 s was used for each
measurement.

In Fig.~\ref{fig:fig3}(b), we plot the value of $g^{(2)}(0)$ as a
function of average input power into the CROW. $g^{(2)}(0)<0.5$ for
all pump powers that we recorded, indicating that we indeed have a
source that is antibunched and dominantly composed of single
photons~\footnote{$N_{AB}$ and $N_{AC}$ are given by the average
detection rates on SPADS B and C multiplied by the integration time,
respectively. $N_{A}$ is the average photon detection rate on SPAD A
mutiplied by the integration time.  The one standard deviation
uncertainties on these values are propagated to generate the error
bars in Fig. 3(a).}.  The minimum value we measured is
$g^{(2)}(0)=0.19\pm0.03$ at $\approx1.7$ mW of average power into
the CROW.  At lower power levels in our experiment, $g^{(2)}(0)$ may
be limited by detector dark counts, while at higher power levels,
the increase in $g^{(2)}(0)$ is likely due to the increased
multi-photon probability as multiple photon pairs are generated in
each optical pulse.  The maximum power levels we can inject into the
CROW were ultimately limited by the damage threshold of the input
couplers. In Fig.~\ref{fig:fig3}(c), we plot the heralding rate
(detection rate of L-band photons by SPAD A) at the CROW output. At
the minimum value of $g^{(2)}(0)$, the heralding rate was
$\approx$220 kHz ($\approx$0.028 photons/pulse). As the input power
to the CROW increases, the generation rate of heralding photons
saturated near 1 MHz due to TPA/FCA effects in silicon. Under pulsed
pumping (2.5 ns pulses, 8 MHz trigger rate) and at the input power
corresponding to the minimum value of $g^{(2)}(0)$, CAR$\approx$15
was measured without dark count subtraction. Subtraction of dark
count coincidences (due to dark counts on both detectors as well
dark counts on one detector and photon detection events on the other
detector) yields CAR=23.8$\pm$5.6.  This significant correction
indicates that $g^{(2)}(0)$ reported in Fig.~\ref{fig:fig3} may
contain a large contribution due to dark counts.

%A significant challenge to quantum optical measurements in the
%telecommunications band is in single photon
%detection~\cite{ref:Ribordy,ref:Hadfield_nphoton_09}.  In comparison
%to Si SPADS used at shorter wavelengths, InGaAs/InP SPADs have lower
%quantum efficiencies, higher dark count rates, and require much
%longer deadtimes after photon detection events in order to
%sufficiently reduce afterpulsing, which strongly limits the rate at
%which they can be triggered. Heralded single photon measurements in
%silicon-based nanophotonic devices are additionally challenging
%because both signal and idler photons are necessarily at wavelengths
%$>$1 $\mu$m, due to silicon's strong linear absorption at shorter
%wavelengths. This then requires all three detectors in the heralded
%$g^{(2)}$ measurement to be InGaAs/InP SPADs. In comparison, sources
%produced in media that are more broadly transparent can be
%engineered to produce very wide signal-idler separation, with signal
%photons in the telecommunications-band and idler photons at $<$1
%$\mu$m~\cite{ref:Fasel_Zbinden_NJP}.  The measurements performed in
%this work were enabled by recent advances in InGaAs/InP
%SPADs~\cite{ref:Zappa_Tosi_Cova} that allow them to be triggered at
%faster rates than previous devices~\cite{ref:Ribordy} without
%suffering adversely in terms of dark counts or afterpulsing.  This
%allowed us to operate the heralding detector (SPAD A) at a
%relatively fast rate, keeping the overall measurement time
%reasonable.

In summary, we have demonstrated a telecommunications-band silicon
heralded single photon source.  Spontaneous four-wave-mixing in a
35-ring silicon coupled resonator optical waveguide generated photon
pairs spaced by 40 nm, with a coincidence-to-accidental ratio $>10$
for continuous wave pumping and $>20$ for pulsed pumping.  Three
InGaAs/InP single photon counters were used to perform a measurement
in which the detection of the idler photons from the pairs triggers
a photon correlation measurement on the corresponding signal
photons. We measured antibunching with $g^{(2)}(0)$=0.19$\pm$0.03,
indicating a source that is dominantly composed of single photons.
Our demonstration of heralded single photon generation within a
silicon photonics platform, for which sophisticated levels of
switching and multiplexing have been shown~\cite{ref:Soref_JSTQE},
is a step towards integration of multiple heralded sources to create
quasi-deterministic single photon
sources~\cite{ref:Migdall_PRA,ref:Pittman_Franson_PRA,ref:Ma_Zeilinger_PRA}
that may then be combined with waveguide quantum photonic
circuits~\cite{ref:Politi_OBrien} and single photon
counters~\cite{ref:Pernice_Tang_arXiv} to achieve high levels of
functionality in future quantum information processing applications.

This work was supported by the National Science Foundation under
grants ECCS-0642603, ECCS-0925399, NSF-GOALI collaboration with IBM,
NSF-NIST supplement, UCSD-Calit2 and the CNST/Maryland Nanocenter
cooperative agreement. J. R. Ong acknowledges support from Agency
for Science, Technology and Research (A{*}STAR) Singapore. We thank
Nick Bertone from Optoelectronic Components for his help in setting
up this collaboration.

\end{document}